\documentclass[showpacs, showkeys,twocolumn]{revtex4}
\begin{document}
\title{ Quark-Antiquark and Diquark
Condensates in Vacuum in a 3D Two-Flavor Gross-Neveu
Model\footnote{The project supported by the National Natural
Science Foundation of China under Grant No.10475113.}}
\author{Zhou Bang-Rong }
\affiliation{College of Physical Sciences, Graduate School of the
Chinese Academy of Sciences, Beijing 100049, China}\affiliation{
CCAST (World Laboratory), P.O.Box 8730, Beijing 100080, China}
\date{}
\begin{abstract}
The effective potential analysis indicates that, in a 3D
two-flavor Gross-Neveu model in vacuum, depending on less or
bigger than the critical value 2/3 of $G_S/H_P$, where $G_S$ and
$H_P$ are respectively the coupling constants of scalar
quark-antiquark channel and pseudoscalar diquark channel, the
system will have the ground state with pure diquark condensates or
with pure quark-antiquark condensates, but no the one with
coexistence of the two forms of condensates. The similarities and
differences in the interplay between the quark-antiquark and the
diquark condensates in vacuum in the 2D, 3D and 4D two-flavor
four-fermion interaction models are summarized. 
\end{abstract}
\maketitle
\section{Introduction\label{Intro}}
It has been shown by effective potential approach that in a
two-flavor 4D Nambu-Jona-Lasinio (NJL) model \cite{kn:1}, even
when temperature $T=0$ and quark chemical potential $\mu=0$, i.e.
in vacuum, there could exist mutual competition between the
quark-antiquark condensates and the diquark condensates
\cite{kn:2}.  Similar situation has also emerged from a 2D
two-flavor Gross-Neveu (GN) model \cite{kn:3} except some
difference in the details of the results \cite{kn:4}. An
interesting question is that if such mutual competition between
the two forms of condensates is a general characteristic of this
kind of two-flavor four-fermion interaction models? For answer to
this question, on the basis of research on the 4D NJL model and
the 2D GN model, we will continue to examine a 3D two-flavor GN
model in similar way. The results will certainly deepen our
understanding
of the feature of the four-fermion interaction models. \\
\indent We will use the effective potential in the mean field
approximation which is equivalent to the leading order of $1/N$
expansion.  It is indicated that a 3D GN model is renormalizable
in $1/N$ expansion \cite{kn:5}.
\section{Model and its symmetries}
The Lagrangian of the model will be expressed by
\begin{eqnarray}
{\cal L}&=&\bar{q}i\gamma^{\mu}\partial_{\mu} q
+G_S[(\bar{q}q)^2+(\bar{q}\vec{\tau}q)^2]\nonumber\\
&&+H_P\sum_{A=2,5,7}(\bar{q}\tau_2\lambda_Aq^C)
    ({\bar{q}}^C\tau_2\lambda_Aq).
\end{eqnarray}
All the denotations used in Eq.(1) are the same as the ones in the
2D GN model given in Ref.\cite{kn:4}, except that the dimension of
space-time is changed from 2 to 3 and the coupling constant $H_S$
of scalar diquark interaction channel is replaced by the coupling
constant $H_P$ of pseudoscalar diquark interaction channel. Now
the matrices $\gamma^{\mu} (\mu=0,1,2)$ and the charge conjugate
matrix $C$ are taken to be $2\times 2$ ones and have the explicit
forms
\begin{equation}
\gamma^0=\left(
\begin{array}{cc}
  1 & 0 \\
  0 & -1 \\
\end{array}
\right),\;\gamma^1=\left(
\begin{array}{cc}
  0 & i \\
  i & 0 \\
\end{array}
\right),\;\gamma^2=\left(
\begin{array}{cc}
  0 & 1 \\
  -1 & 0 \\
\end{array}
\right)=C.
\end{equation}
It is emphasized that, in 3D case, no "$\gamma_5$" matrix can be
defined, hence the third term in the right-handed side of Eq.(1)
will be the only possible color-anti-triplet diquark interaction
channel which could lead to Lorentz-invariant diquark condensates,
where we note that the matrix $C\tau_2\lambda_A$ is antisymmetric.
Without "$\gamma_5$", the Lagrangian (1) will have no chiral
symmetry. Except this, it is not difficult to verify that the
symmetries of $\mathcal{L}$ include:
\begin{enumerate}
\item continuous flavor and color symmetries $SU_f(2)\otimes
SU_c(3)\otimes U_f(1)$; \item discrete symmetry R: $q\rightarrow
-q$; \item parity $\mathcal{P}$: $q(t,\vec{x})\rightarrow
\gamma^0q(t,-\vec{x})$ and $q^C(t,\vec{x})\rightarrow
-\gamma^0q^C(t,-\vec{x})$; \item time reversal $\mathcal{T}$:
$q(t,\vec{x})\rightarrow \gamma^2q(-t,\vec{x})$ and
$q^C(t,\vec{x})\rightarrow -\gamma^2q^C(-t,\vec{x})$; \item charge
conjugate $\mathcal{C}$: $q\leftrightarrow q^C$; \item special
parity $\mathcal{P}_1$: $q(t, x^1, x^2) \rightarrow \gamma^1 q(t,
-x^1, x^2)$ and $q^C(t, x^1, x^2) \rightarrow -\gamma^1 q(t, -x^1,
x^2)$; \item special parity $\mathcal{P}_2$: $q(t, x^1, x^2)
\rightarrow \gamma^2 q(t, x^1, -x^2)$ and $q^C(t, x^1, x^2)
\rightarrow -\gamma^2 q^C(t, x^1, -x^2)$.
\end{enumerate}
 If the quark-antiquark condensates $\langle\bar{q}q\rangle$ could be
formed, then the time reversal $\mathcal{T}$, the special parities
$\mathcal{P}_1$ and $\mathcal{P}_2$ will be spontaneously broken
\cite{kn:6}. If the diquark condensates
$\langle\bar{q}^C\tau_2\lambda_2q\rangle$ could be formed, then
the color symmetry $SU_c(3)$ will be spontaneously broken down to
$SU_c(2)$ and the flavor number $U_f(1)$ will be spontaneously
broken but a "rotated" electric charge $U_{\tilde{Q}}(1)$ and a
"rotated" quark number $U^{\prime}_{q}(1)$ leave unbroken
\cite{kn:7}. In addition, the parity $\mathcal{P}$ will be
spontaneously broken, though all the other discrete symmetries
survive. This implies that the diquark condensates
$\langle\bar{q}^C\tau_2\lambda_2q\rangle$ will be a pseudoscalar.
In this paper we will neglect discussions of the Goldstone bosons
induced by breakdown of the continuous symmetries and pay our main
attention to the problem of interplay between the above two forms
of condensates.
\section{Effective potential in mean field approximation}
Define the order parameters in the 3D GN model by
\begin{equation}
\sigma = -2G_S\langle\bar{q}q\rangle \;\mathrm{and} \;
 \Delta = -2H_P\langle\bar{q}^C\tau_2\lambda_2q\rangle,
\end{equation}
then in the mean field approximation, the Lagrangian (1) can be
rewritten by
\begin{equation}
\mathcal{L}=\bar{\Psi}(x)S^{-1}(x)\Psi(x)-\frac{\sigma^2}{4G_S}-\frac{|\Delta|^2}{4H_P},
\end{equation}
where
$$\Psi(x)=\frac{1}{\sqrt{2}}\left(
\begin{array}{c}
  q(x)\\
  q^C(x) \\
\end{array}
\right)\; \mathrm{and}\; \bar{\Psi}(x)=\frac{1}{\sqrt{2}}\left(
\begin{array}{cc}
  \bar{q}(x) & \bar{q}^C(x) \\
\end{array}
\right)
$$
are the expressions of the quark fields in the Nambu-Gorkov basis
\cite{kn:8}. In the momentum space, the inverse propagator
$S^{-1}(x)$ for the quark fields may be expressed by
\begin{equation}
S^{-1}(p)=\left(
\begin{array}{cc}
  \not\!{p}-\sigma & -\tau_2\lambda_2\Delta \\
  -\tau_2\lambda_2\Delta^* & \not\!{p}-\sigma \\
\end{array}
\right),\; \not\!{p}=\gamma^{\mu}p_{\mu}.
\end{equation}
The effective potential corresponding to $\mathcal{L}$ given by
Eq.(4) becomes
\begin{equation}
V(\sigma, |\Delta|)=\frac{\sigma^2}{4G_S}+\frac{|\Delta|^2}{4H_P}
+i\int\frac{d^3p}{(2\pi)^3}\frac{1}{2}\mathrm{Tr}\ln
  S^{-1}(p)S_0(p).
\end{equation}
Similar to the case of the 2D NG model \cite{kn:4}, the
calculations of $\mathrm{Tr}$ for (red, green) and blue color
degrees of freedom can be made separately thus Eq.(6) will be
reduced to
\begin{widetext}
\begin{equation}
V(\sigma,
|\Delta|)=\frac{\sigma^2}{4G_S}+\frac{|\Delta|^2}{4H_P}+2i
\int\frac{d^3p}{(2\pi)^3} \left[\ln
\frac{p^2-(\sigma-|\Delta|)^2+i\varepsilon}{p^2+i\varepsilon}+\ln
\frac{p^2-(\sigma+|\Delta|)^2+i\varepsilon}{p^2+i\varepsilon} +\ln
\frac{p^2-\sigma^2+i\varepsilon}{p^2+i\varepsilon}\right]
\end{equation}
\end{widetext}
After the Wick rotation, we may define and calculate in 3D
Euclidean momentum space
\begin{eqnarray}
 I(a^2)&=&\int\frac{d^3\bar{p}}{(2\pi)^3}\ln\frac{\bar{p}^2+a^2}{\bar{p}^2} \nonumber \\
  &=&\frac{1}{2\pi^2}\left(a^2\Lambda-\frac{2}{3}a^3\arctan\frac{\Lambda}{a}\right)\nonumber \\
  &\simeq&\frac{1}{2\pi^2}\left(a^2\Lambda-\frac{\pi}{3}|a|^3\right),
  \; \mathrm{if}\; \Lambda\gg |a|,
\end{eqnarray}
where $\Lambda$ is the 3D Euclidean momentum cut-off. Assume that
$\Lambda\gg |\sigma-|\Delta||$, $\Lambda\gg \sigma+|\Delta|$ and
$\Lambda\gg \sigma$, then by means of Eq.(8) we will obtain the
final expression of the effective potential in the 3D GN model
\begin{eqnarray}
 V(\sigma, |\Delta|)&=& \frac{\sigma^2}{4G_S}+\frac{|\Delta|^2}{4H_P}
 -\frac{1}{\pi^2}(3\sigma^2+2|\Delta|^2)\Lambda\nonumber \\
 &&+\frac{1}{3\pi}\left[6\sigma^2|\Delta|+2|\Delta|^3+\sigma^3\right.\nonumber \\
 &&\left.+ 2\theta(\sigma-|\Delta|)(\sigma-|\Delta|)^3\right].
\end{eqnarray}
\\
\section{Ground states}
Equation (9) provide the possibility to discuss the ground states
of the model analytically. The extreme value conditions $\partial
V(\sigma, |\Delta|)/\partial \sigma=0$ and $\partial V(\sigma,
|\Delta|)/\partial |\Delta|=0$ will lead to the equations
\begin{eqnarray}
\sigma\left(\frac{1}{2G_S}-\frac{6\Lambda}{\pi^2}
+\frac{4|\Delta|}{\pi}+\frac{\sigma}{\pi}\right)&&\nonumber\\
+\frac{2}{\pi}\theta(\sigma-|\Delta|)(\sigma-|\Delta|)^2&=&0,
\end{eqnarray}
\begin{eqnarray}
|\Delta|\left(\frac{1}{2H_P}-\frac{4\Lambda}{\pi^2}+
\frac{2|\Delta|}{\pi}\right)&&\nonumber\\
+\frac{2}{\pi}[\sigma^2-\theta(\sigma-|\Delta|)(\sigma-|\Delta|)^2]&=&0.
\end{eqnarray}
Define the expressions
$$
K=\left|\begin{array}{cc}
  A & B \\
  B & C \\
\end{array}\right|=AC-B^2,
$$
where $A$, $B$ and $C$ represent the second order derivatives of
$V(\sigma, |\Delta|)$ with the explicit expressions
\begin{widetext}
\begin{eqnarray}
A&=&\frac{\partial^2V}{\partial\sigma^2}=\frac{1}{2G_S}-\frac{6\Lambda}{\pi^2}
+\frac{4|\Delta|}{\pi}+\frac{2\sigma}{\pi}
+\frac{4}{\pi}\theta(\sigma-|\Delta|)(\sigma-|\Delta|),
  \nonumber \\
B&=&\frac{\partial^2V}{\partial\sigma\partial|\Delta|}=
\frac{\partial^2V}{\partial|\Delta|\partial\sigma}=
\frac{4}{\pi}[\sigma-\theta(\sigma-|\Delta|)(\sigma-|\Delta|)],\nonumber \\
C&=&\frac{\partial^2V}{\partial|\Delta|^2}=
\frac{1}{2H_P}-\frac{4\Lambda}{\pi^2} +\frac{4|\Delta|}{\pi}
+\frac{4}{\pi}\theta(\sigma-|\Delta|)(\sigma-|\Delta|).
\end{eqnarray}
\end{widetext}
Equations (10) and (11) have the four different solutions which
will be discussed in proper order as follows.\\
(i) ($\sigma, |\Delta|$)=(0,0). It is a maximum point of
$V(\sigma, |\Delta|)$, since in this case we have
$$
A=\frac{1}{2G_S}-\frac{6\Lambda}{\pi^2}<0 \; \;\mathrm{and}\;\;
K=A\left(\frac{1}{2H_P}-\frac{4\Lambda}{\pi^2}\right)>0,
$$
assuming Eqs. (10) and (11) have solutions of non-zero $\sigma$
and
$|\Delta|$. \\
(ii) ($\sigma, |\Delta|$)=($\sigma_1$,0), where the non-zero
$\sigma_1$ satisfies the equation
\begin{equation}
\frac{1}{2G_S}-\frac{6\Lambda}{\pi^2}+\frac{3\sigma_1}{\pi}=0.
\end{equation}
When Eq. (13) is used, we obtain
\begin{eqnarray*}
  A &=&\frac{3\sigma_1}{\pi}, \\
  K=&=&A\left(\frac{1}{2H_P}-\frac{1}{3G_S}+\frac{2\sigma_1}{\pi}\right)>0,
  \;\; \mathrm{if}\;\; \frac{G_S}{H_P}>\frac{2}{3}.
\end{eqnarray*}
Hence ($\sigma_1$,0) will be a minimum point of $V(\sigma,
|\Delta|)$ when $G_S/H_P>2/3$.\\
(iii) ($\sigma$, $|\Delta|$)= (0, $\Delta_1$), where non-zero
$\Delta_1$ obeys the equation
\begin{equation}
\frac{1}{2H_P}-\frac{4\Lambda}{\pi^2}+\frac{2\Delta_1}{\pi}=0.
\end{equation}
By using Eq.(14) we may get
$$
  A =\frac{1}{2G_S}-\frac{3}{4H_P}+\frac{\Delta_1}{\pi},\;\;
  K =A\frac{2\Delta_1}{\pi}.
$$
Obviously, (0,$\Delta_1$) will be a minimum point of $V(\sigma,
|\Delta|)$ when $G_S/H_P<2/3$.\\
(iv) ($\sigma, |\Delta|$)=($\sigma_2, \Delta_2$). In view of
existence of the function $\theta(\sigma-|\Delta|)$ in Eqs.(10)
and (11), we have to consider the case of $\sigma_2>\Delta_2$ and
$\sigma_2<\Delta_2$ respectively.\\
\indent (a) $\sigma_2>\Delta_2$. In this case, Eqs.(10) and (11)
will become
$$
\sigma_2\left(\frac{1}{2G_S}-\frac{6\Lambda}{\pi^2}\right)
+\frac{1}{\pi}(3\sigma_2^2+2\Delta_2^2)=0,
$$ $$
\frac{1}{2H_P}-\frac{4\Lambda}{\pi^2}+\frac{4\sigma_2}{\pi}=0.
$$
From them we can get
$$
A=\frac{3\sigma_2^2-2\Delta_2^2}{\pi\sigma_2}>0,\;\;
K=-\frac{16\Delta_2^2}{\pi^2}<0.
$$
Thus it is turned out that ($\sigma_2$, $\Delta_2$) will be
neither a maximum nor a minimum point of $V(\sigma, |\Delta|)$ if
$\sigma_2>\Delta_2$.\\
\indent (b) $\sigma_2<\Delta_2$. Now Eqs. (10) and (11) are
changed into
\begin{equation}
\frac{1}{2G_S}-\frac{6\Lambda}{\pi^2}
+\frac{4\Delta_2+\sigma_2}{\pi}=0,
\end{equation}
\begin{equation}
\Delta_2\left(\frac{1}{2H_P}-\frac{4\Lambda}{\pi^2}\right)+
2\frac{\sigma_2^2+\Delta_2^2}{\pi}=0.
\end{equation}
Hence we will have the results that
$$
A=\frac{\sigma_2}{\pi},\;
K=\frac{2\sigma_2}{\pi^2\Delta_2}(\Delta_2^2-\sigma_2^2-8\sigma_2\Delta_2),
$$
from which it may be deduced that only if
\begin{equation}
 \sigma_2<(\sqrt{17}-4)\Delta_2,
\end{equation}
$(\sigma_2, \Delta_2)$ is just a minimum point of
$V(\sigma,|\Delta|)$. On the other hand, from Eqs. (15) and (16)
obeyed by $\sigma_2$ and $\Delta_2$ we may get
\begin{eqnarray}
\frac{1}{2H_P}-\frac{1}{3G_S}&=&\frac{2}{\pi\Delta_2}
\left(\frac{\sqrt{13}-1}{6}\Delta_2+\sigma_2\right)\nonumber\\
&&\times\left(\frac{\sqrt{13}+1}{6}\Delta_2-\sigma_2\right).
\end{eqnarray}
Equation (18) indicates that for the minimum point $(\sigma_2,
\Delta_2)$ satisfying Eq.(17) one will certainly have
$G_S/H_P>2/3$. Taking this and the result obtained in case (ii)
into account we see that if $G_S/H_P>2/3$ the effective potential
$V(\sigma, |\Delta|)$ will have two possible minimum points
$(\sigma_1,0)$ and $(\sigma_2, \Delta_2)$.  To determine which one
of the two minimum points is the least value point of $V$, we must
make a comparison between $V(\sigma_1,0)$ and
$V(\sigma_2,\Delta_2)$ with the constraint given by Eq.(17). In
fact, it is easy to find out that when Eq.(13) is used,
\begin{equation}
V(\sigma_1,0)=-\frac{\sigma_1^3}{2\pi},
\end{equation}
and that when Eqs. (15) and (16) are used,
\begin{equation}
V(\sigma_2,\Delta_2)=-\frac{1}{3\pi}\left(\Delta_2^3
+3\sigma_2^2\Delta_2+\frac{\sigma_2^3}{2}\right).
\end{equation}
By comparing Eq.(13) with Eq.(15) we may obtain the relation
\begin{equation}
3\sigma_1=\sigma_2+4\Delta_2.
\end{equation}
By means of Eqs.(19)-(21) it is easy to verify that
\begin{eqnarray*}
  V(\sigma_1,0)-V(\sigma_2,\Delta_2)&=&-\frac{1}{3\pi}
  \left\{\frac{1}{9}(23\Delta_2^3-4\sigma_2^3)\right.\\
  &&\left.+\frac{\sigma_2\Delta_2}{3}(8\Delta_2-7\sigma_2)\right\}<0,
\end{eqnarray*}
when Eq.(17) is satisfied.  This result indicates that when
$G_S/H_P>2/3$, the least value point of $V(\sigma, |\Delta|)$ will
be $(\sigma_1, 0)$ but not $(\sigma_2,\Delta_2)$. \\ \\
\indent In summary, if the necessary conditions
$G_S\Lambda>\pi^2/12$ and $H_P\Lambda > \pi^2/8$ for non-zero
$\sigma$ and $\Delta$ are satisfied, then the least value points
of the effective potential $V(\sigma, |\Delta|)$ will be at
\begin{equation}
(\sigma, |\Delta|)=\left\{\begin{array}{c}
  (0,\Delta_1) \\
  (\sigma_1,0) \\
\end{array}\right.\; \mathrm{if}\; \left\{\begin{array}{cc}
  0\leq & G_S/H_P<2/3 \\
        &G_S/H_P>2/3 \\
\end{array}\right..
\end{equation}
As a result, in the ground state of the 3D two-flavor GN model,
depending on that the ratio $G_S/H_P$ is either bigger or less
than 2/3, one will have either pure quark-antiquark condensates or
pure diquark condensates, but no coexistence of the two forms of
condensates could happen.
\section{Concluding remarks}
The result (22) in the 3D GN model can be compared with the ones
in the 4D NJL model and in the 2D GN model. The minimal points of
the effective potential $V(\sigma,|\Delta|)$ for the latter models
have been obtained and are located respectively at
\begin{widetext}
\begin{equation}
(\sigma,|\Delta|)=\left\{\begin{array}{lc}
  (0, & \Delta_1) \\
  (\sigma_2, & \Delta_2) \\
  (\sigma_1, & 0) \\
\end{array}\right.\;\;\mbox{if}\;\;
\left\{\begin{array}{cccl}
  &0&\leq&G_S/H_S<2/[3(1+C)] \\
  &2/[3(1+C)]&<&G_S/H_S<2/3 \\
  &&&G_S/H_S>2/3 \\
\end{array}\right.
\end{equation}
\end{widetext}
with $C=(2H_S\Lambda_4^2/\pi^2-1)/3$ and $\Lambda_4$ denoting the
4D Euclidean momentum cutoff in the 4D two-flavor NJL model, if
the necessary conditions $G_S\Lambda_4^2>\pi^2/3$ and
$H_S\Lambda_4^2> \pi^2/2$ for non-zero $\sigma$ and $\Delta$ are
satisfied \cite{kn:2}, and
\begin{equation}
(\sigma,|\Delta|)=\left\{\begin{array}{lc}
  (0, & \Delta_1) \\
  (\sigma_2, & \Delta_2) \\
  (\sigma_1, & 0) \\
\end{array}\right.\;\;\mbox{if}\;\;
\left\{\begin{array}{c}
  G_S/H_S=0 \\
  0<G_S/H_S<2/3 \\
  G_S/H_S>2/3 \\
\end{array}\right.
\end{equation}
in the 2D two-flavor GN model \cite{kn:4}. In Eqs.(23) and (24),
$G_S$ and $H_S$ always represent the coupling constants in scalar
quark-antiquark channel and scalar
diquark channel separately. \\
\indent By a comparison among Eqs.(22)-(24) it may be found that
the three models lead to very similar results. In all the three
models, the interplay between the quark-antiquark and the diquark
condensates in vacuum depends on the ratio $G_S/H_D$ ($D=S$ for
the 4D and 2D model and $D=P$ for the 3D model). In particular,
the diquark condensates could emerge (in separate or coexistent
pattern) only if $G_S/H_D<2/3$. This is probably a general
characteristic of the considered two-flavor four-fermion models,
since in these models the color number of the quarks participating
in the diquark condensates and in the quark-antiquark condensates
is just 2 and 3 respectively. However, there are also some
differences in the pattern realizing the diquark condensates among
the three models, though the pure quark-antiquark condensates
arise only if $G_S/H_D>2/3$ in all of them. In the 2D GN model,
the pure diquark condensates emerge only if $G_S/H_S=0$ and this
is different from the 4D NJL model where the pure diquark
condensates may arise if $G_S/H_S$ is in a finite region below
2/3. Another difference is that in the 3D GN model, there is no
coexistence of the quark-antuquark condensates and the diquark
condensates but such coexistence is clearly displayed in the 4D
and 2D model. This implies that in the 3D GN model, $G_S/H_P=2/3$
becomes the critical value which distinguishes between the ground
states with the pure diquark condensates and with the pure
quark-antiquark
condensates.\\
\indent It is also indicated that if the two-flavor four-fermion
interaction models are assumed to be simulations of QCD (of
course, only the 4D NJL model is just the true one) and the
four-fermion interactions are supposed to come from the heavy
color gluon exchange interactions
$-g(\bar{q}\gamma^{\mu}\lambda_aq)^2\;(a=1,\cdots,8; \mu=0,
\cdots, D-1)$ via the Fierz transformation \cite{kn:7}, then one
will find that in all the three models, for the case of two
flavors and three colors the ratio $G_S/H_D$ are always equal to
4/3 which is larger than the above critical value 2/3. From this
we can conclude that there will be only the pure quark-antiquark
condensates and no diquark condensates in the ground states of all
these models in vacuum.

\end{document}